\documentclass[twocolumn,amsmath,amssymb,prb]{revtex4}
\usepackage{psfig}

\def\AmS{{\protect\the\textfont2
        A\kern-.1667em\lower.5ex\hbox{M}\kern-.125emS}}

\makeatletter
\def\thepage{1-\@arabic\c@page}
\def\@pnumwidth{2em}
\usepackage{graphicx}
\begin{document}
   

\title{Exchange interactions and Curie temperature in (GaMn)As}

\author{
L.M. Sandratskii\thanks{lsandr@mpi-halle.de} and P. Bruno}

\affiliation{Max-Planck Institut f\"ur Mikrostrukturphysik, D-06120 Halle, Germany }
 
\begin{abstract}
We use supercell and frozen-magnon approaches to study the dependence 
of the magnetic interactions in (Ga,Mn)As on the Mn concentration. We report 
the parameters of the exchange interaction between Mn spins and the 
estimates of the Curie temperature within the mean-field and random-phase
approximations. In agreement with
experiment we obtain a nonmonotonous dependence of the Curie temperature
on the Mn concentration. We estimate the dependence of the Curie temperature
on the concentration of the carries in the system and show that the decrease
of the number of holes in the valence band leads to fast decrease of the 
Curie temperature. We show that the hole states of the valence band are
more efficient in mediating the exchange interaction between Mn spins than
the electron states of the conduction band.
\end{abstract}   
\maketitle

\section{Introduction}

An important current problem on the way to the practical use
of the spin-transport in semiconductor devices is the design of the
materials that make possible the injection of the spin-polarized electrons 
into semiconductor at room temperature. One of the promising
classes of materials are the diluted magnetic semiconductors (DMS) of the III-V type. A strong 
interest to these systems were attracted by the
observation of the ferromagnetism in (GaMn)In \cite{ohno_gain} 
and (GaMn)As \cite{ohno_gaas} with the Curie temperature 
of Ga$_{0.947}$Mn$_{0.053}$As as high as 110 K. To design the material with Curie
temperature higher than room temperature the knowledge of physical mechanisms governing
the exchange interactions in these systems is of primary importance.

The theoretical works on the ferromagnetism in the DMS systems can
be separated into two groups. First group models the problem with an effective Hamiltonian
containing experimentally determined parameters. This part of the
studies is recently reviewed in Refs. \cite{dietl_rev,dms_mac}. This 
paper belongs to the second group of studies that  
are based on the parameter-free calculations within the density
functional theory. Several calculations have been performed recently along this
line. In Refs. \cite{akai,sato_rev,korz02},
the coherent potential approximation (CPA) was used to study the magnetic structure,
density of states, total energy, and chemical trends in (III,Mn)V. 
In Refs. \cite{zhao,sanv_prb01,sanv_apl01,mark01}, the calculations were
performed for a series of magnetic states of (III,Mn)V using
supercells of the zinc-blende structure and focusing basically on the same 
physical quantities as the CPA studies. 
Since the CPA is a single-site
theory which neglects any short range order in the disordered subsystems
the CPA and supercell approaches are 
complimentary. In Refs. \cite{akai,sanv_apl01,korz02}, the influence of the 
antisite defects on the magnetic properties of the DMS is discussed.
In Ref. \cite{mark01}, the total energy of various collinear magnetic
configurations are used to estimate the parameters of the exchange interaction
between the 3d atoms forming clusters in III-V semiconductors.    

The purpose of the present paper is parameter-free calculation
of the exchange interactions and Curie temperature in (GaMn)As for various
concentrations of Mn. The study is based on the supercell approach. 

Density functional theory (DFT) has proved to be very successful in 
the parameter-free description of the ground state magnetic properties of complex 
systems (see, e.g. \cite{dreysse}). 
Recently much attention has been devoted to the application of the methods of the DFT
to the studies of the low-energy excitations of magnetic systems and
the magnetic phase transitions \cite{likaan,ulku,rojo,haespe,nikl,savrasov,brniwa,pakutu,gredfa}. 
The Stoner theory, which relates 
the finite temperature effects to the temperature variation of the
Fermi-Dirac distribution appeared to be unable to describe the
temperature properties of the magnetic systems with itinerant electrons. \cite{moriya}
A necessary feature of the theoretical description of the finite
temperature effects is an account for transversal fluctuations of the  
local magnetization. A most consequent method of the calculation
of the low-energy magnetic excitations is based on the evaluation
of the nonuniform and frequency dependent enhanced magnetic
susceptibility. \cite{savrasov} This approach is, however, 
computationally very complicated
and up to now has been successfully applied to the simplest magnetic
systems only. 

A more tractable approach to the study of both the spin-wave excitations
and the thermodynamics of magnetic systems is based on an adiabatic
treatment of the atomic magnetic moments. \cite{likaan,ulku,rojo,haespe,nikl,brniwa,pakutu,gredfa} 
In this approach the
account for noncollinear configurations of atomic moments is essential.
An effective method for the estimation of the parameters
of the interatomic exchange interaction and spin-wave energies is suggested 
by the frozen-magnon approach. \cite{haespe,gredfa} 
This approach is based on the total energy calculation for
spiral magnetic configurations. Because of the generalized
translational periodicity of spin spirals \cite{herring}
such calculations can be performed very efficiently. \cite{sand86}
Additional help is provided by the force theorem \cite{likaan} that allows to use
band energy of non-selfconsistent frozen magnon states
for the estimation of the total-energy differences.
In the present paper this approach is used to study the exchange interactions 
and Curie temperature in $\rm Ga_{1-x}Mn_xAs$ with various
concentrations of the Mn impurities. 

\section{Calculational scheme}
\label{sec:calc_scheme}

The calculations are based on the supercell approach where one of the 
Ga atoms in a supercell of zinc-blende GaAs is replaced by the Mn atom. The concentration of 
Mn depends on the size of the super cell. The following concentrations $x$
have been studied in this work: 0.25, 0.125, 0.0625, and 0.03125. For comparison,
the calculation of the density of states (DOS) of pure MnAs and GaAs are also
performed. In oder to better identify trends we investigate here a
concentration range that is much wider the one accessible to experiment sofar.

The calculations were carried out with the augmented spherical waves (ASW) 
method. \cite{wikuge}
In all calculations the lattice parameter 
was chosen to be equal to the experimental lattice parameter of GaAs.
Two empty spheres per formula unit have been used in the calculations.
The positions of empty spheres are (0.5, 0.5, 0.5) and (0.75, 0.75, 0.75).
Radii of all atomic spheres were chosen to be equal. 
Depending on the concentration of Mn, the super cell is cubic
(x=25\%, 
$a\times a\times a$, and x=3.125\%, $2a\times 2a\times 2a$) 
or tetragonal 
(x=12.5\%, 
$a\times a \times 2a$ and 6.25\%, $2a\times 2a\times a$).
 
To describe the exchange interactions in the system
we use an effective Heisenberg
Hamiltonian of classical spins
\begin{equation}
\label{eq:hamiltonian}
H_{eff}=-\sum_{i\ne j} J_{ij} {\bf e}_i{\bf e}_j
\end{equation}
where $J_{ij}$ is an exchange interaction between two Mn sites $(i,j)$
and ${\bf e}_i$ is the unit vector pointing in the direction 
of the magnetic moments at site $i$. 

To estimate the parameters of the Mn-Mn exchange interaction we
performed calculation for the following frozen-magnon configurations:
\begin{equation}
\theta_i=const, \:\: \phi_i={\bf q \cdot R}_i
\end{equation}
where $\theta_i$ and $\phi_i$ are the polar and azimuthal angles of vector
${\bf e}_i$, ${\bf R}_i$ is the position of the $i$th Mn atom. 
The directions of the induced moments in the atomic spheres of Ga and As
and in the empty spheres were kept to be parallel to the $z$ axis. 

It can be shown that within the Heisenberg model~(\ref{eq:hamiltonian})
the energy of such configurations can be represented in the form
\begin{equation}
\label{eq:e_of_q}
E(\theta,{\bf q})=E_0(\theta)-\frac{\theta^2}{2} J({\bf q})
\end{equation}
where $E_0$ does not depend on {\bf q} and $J({\bf q})$ is the 
Fourier transform of the parameters of the exchange interaction between 
pairs of Mn atoms:
\begin{equation}
\label{eq:J_q}
J({\bf q})=\sum_{j\ne0} J_{0j}\:\exp(i{\bf q\cdot R_{0j}}).
\end{equation}
In Eq. (\ref{eq:e_of_q}) angle $\theta$ is assumed to be small.
Using $J({\bf q})$ one can estimate the energies of the spin-wave 
excitations:
\begin{equation}
\omega({\bf q})=\frac{4}{M} [J({\bf 0})-J({\bf q})]=
\frac{8}{M\theta^2}(E(\theta,{\bf q})-E(\theta,{\bf 0}))
\end{equation} 
where $M$ is the atomic moment of the Mn atom.
Performing back Fourier transformation we obtain the parameters of 
the exchange interaction between Mn atoms:
\begin{equation}
\label{eq:J_0j}
J_{0j}=\frac{1}{N}\sum_{\bf q} \exp(-i{\bf q\cdot R_{0j}})J({\bf q}).
\end{equation}
The calculation of $E(\theta,{\bf q})$ for different Mn concentrations
has been performed   
for uniform meshes in the first BZ for $\theta=30^\circ$. The symmetry 
of the system was employed to reduce the amount of calculations.
For cubic supercells (Mn concentrations
25\% and 3.125\%) 
the crystal structure possesses 24 point symmetry 
operations, for the tetragonal unit cells (Mn concentrations 12.5\% 
and 
6.25\%) 
the number of point operations is 8. Since the energy 
as a function of {\bf q} is invariant with respect to the 
reversal of {\bf q} the irreducible part of the BZ is $\frac{1}{48}$th and 
$\frac{1}{16}$th
correspondingly for cubic and tetragonal supercells. 

The number and type of the exchange parameters determined in the
back Fourier transformation with Eq. (\ref{eq:J_0j}) are uniquely determined by the
{\bf q} mesh. The parameters obtained in this procedure guarantee that 
J({\bf q}) calculated according to Eq (\ref{eq:J_q}) reproduce exactly the
calculated values of the total energy.
Calculation for a more dense {\bf q} mesh gives additionally the parameters
of the exchange interaction between more distant atoms. 

The Curie temperature was estimated in the mean-field approximation (MFA)
\begin{equation}
\label{eq:Tc_MFA}
k_BT_C^{MFA}=\frac{2}{3}\sum_{j\ne0}J_{0j}=\frac{M}{6\mu_B}\frac{1}{N}\sum_{\bf q}\omega({\bf q})
\end{equation}
and random phase approximation (RPA) 
\begin{equation}
\frac{1}{k_BT_C^{RPA}}=\frac{6\mu_B}{M}\frac{1}{N} \sum_{\bf q} \frac{1}{\omega({\bf q})}
\end{equation}
To evaluate the RPA value of the Curie temperature
the $\omega({\bf q})$ was considered continuous in the {\bf q} space. In a small sphere
with the center at ${\bf q}=0$  the singular function $ \frac{1}{\omega({\bf q})}$ was
approximated by the function $\frac{1}{Dq^2}$ and then replaced by a
continuous function which has the same value and slope at the 
sphere boundary. The difference between the singular and continuous
functions was integrated analytically. The regular function obtained
was integrated numerically.

Since the $T_C^{MFA}$ and $T_C^{RPA}$ are given by the arithmetic and harmonic
averages of the spin-wave energies, $T_C^{MFA}$ is always larger than  $T_C^{RPA}$. \cite{pakutu}
Physically this difference can be explained by an increased role of the 
low-energy excitations in the case of RPA.  
In the case of the ferromagnetic 3d metals the RPA gives, in general, better
agreement with experiment.  \cite{pakutu} (For a recent application of the RPA
within a model-Hamiltonian approach see, e.g., Ref. \cite{bopa}.)

\section{Density of states}
We begin with the discussion of the trends in the variation of the 
\begin{figure}
\caption{The DOS of Ga$_{1-x}$Mn$_{x}$As. The DOS is given per 
unit cell of the zinc-blende crystal structure. The DOS above(below)
the abscissas axis corresponds to the spin-up(down) states. 
 \label{fig:DOS_tot}}
\includegraphics[width=8cm]{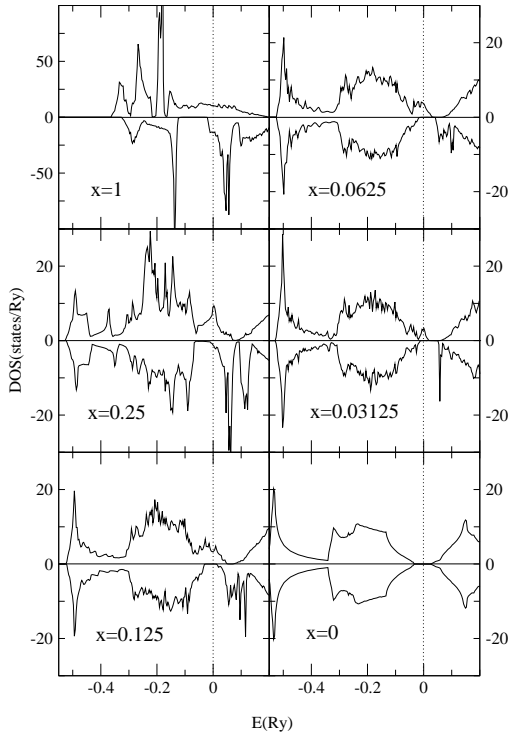}
\end{figure}
electron DOS (Fig.~\ref{fig:DOS_tot}).
For all Mn concentrations studied there is an energy gap 
in the spin-down DOS in an energy interval containing the Fermi level
or close to it.
This gap is about 0.1 Ry in MnAs and decreases to the value
of about 0.05 Ry for lower Mn content. In MnAs and Ga$_{0.75}$Mn$_{0.25}$As, 
the Fermi level is slightly above the gap. For MnAs, the
energy distance between the Fermi level and upper edge of the gap
is about 0.015 Ry. For $x=0.25$, this distance is less than 0.005 Ry.
For lower Mn concentrations the Fermi level lies within the gap
moving from the upper
to the lower part of the gap with decreasing $x$.
This means that for concentrations of Mn less than 25\%  
the calculated ground state is half-metallic. This property is very important
for the efficient spin-injection into semiconductor. \cite{schmidt}
In the half-metallic ferromagnetic state the value of the magnetic moment 
per supercell is integer (Table~\ref{tab:moments}).
\begin{table}
\caption{Magnetic moments in Ga$_{1-x}$Mn$_{x}$As. There are shown the
Mn moment, the induced moment on the nearest As atoms, and the 
magnetic moment of the unit cell (supercell in the case of $x\ne1$).
All moments are in units of $\mu_B$. 
\label{tab:moments}}
\begin{tabular}{lccccc}
&\multicolumn{5}{c}{x}\\
&1&0.25&0.125&0.0625&0.03125\\
\colrule
Mn& 3.76&    3.85&      3.88&          3.94&       3.95\\
As&-0.18 &   -0.046 &  -0.046 &     -0.036 &     -0.032\\
cell&  3.65   &     3.98 &        4.00&         4.00&      4.00\\
\end{tabular}
\end{table} 

The situation with spin-up DOS is different.
In MnAs there is no energy gap close to the Fermi level. For $x$=0.25 a small
gap appears at the energy 0.07 Ry above the Fermi level. With decreasing
$x$, this gap increases and its lower edge becomes closer to the Fermi level. 
At the Mn concentration of 6.25\% 
there is an overlap of the spin-up
and spin-down gaps and an energy gap appears in the total DOS.
The gap in the total DOS can be treated as a gap between the valence and 
conduction bands of the system. 
The upper part of the valence band is not occupied and contains holes.
The hole states are of the spin-up type.

Comparison of our DOS with the corresponding DOS available in the 
scientific literature shows good agreement.
Thus, our DOS for the Mn concentration of 3.125\% 
is close to the corresponding
DOS from Ref. \cite{sanv_prb01} calculated with the use of pseudopotential approach.
Also the partial DOS for x=6.25\% 
presented in Ref.\cite{zhao} is in good agreement
with the curves obtained in our calculations. 
\begin{figure}
\caption{The partial Mn-DOS for Ga$_{1-x}$Mn$_{x}$As. The DOS is given per 
Mn atom.
\label{fig:Mn_DOS}}
\includegraphics[width=8cm]{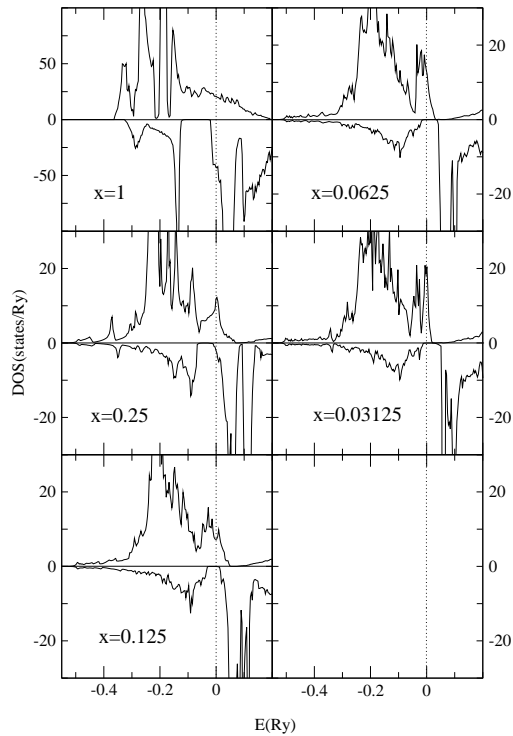}
\end{figure}

The values of the calculated moments in the Mn atomic sphere are collected in 
Table~\ref{tab:moments} and are close to 4$\mu_B$. Similar values have been 
obtained in other calculations within the density functional theory (see, e.g.,
Refs. \cite{zhao,sanv_prb01,mark01}). Taking as an example the system with
$x=3.125\%$ we find that the contribution of the 3d electrons into the 
Mn moment is 3.83$\mu_B$ with the rest 0.12$\mu_B$ coming from the
4s and 4p electrons. Note that the total number of the 3d electrons
in the Mn sphere is $5.30$. The difference between the number of the 3d electrons 
and their contribution into the spin moment results from the presence of $0.73$
spin-down 3d electrons in the Mn atomic sphere (see the spin-down Mn-DOS in
Fig.~\ref{fig:Mn_DOS}). The hybridization between the Mn 3d states and the
states of the valence band of GaAs is crucial for the appearance of the 
occupied spin-down 3d states. In some model-Hamiltonian studies a physical picture is used
which considers the  Mn 3d electrons as strongly localized and forming an atomic spin 
of $S=\frac{5}{2}$. The 
density-functional-theory calculations show that this picture,
although useful in qualitative studies, does not take into account some 
important features of the Mn 3d states. 

\begin{figure}
\caption{Hole distribution and induced moments (in $\mu_B$) for Ga$_{1-x}$Mn$_{x}$As
with $x=0.03125\%$.
All values are given for the coordination spheres. The numbers of atoms in the coordination spheres
are given at the top of the picture.
\label{fig:hole_moment_distribution}}
\includegraphics[width=8cm,angle=-90]{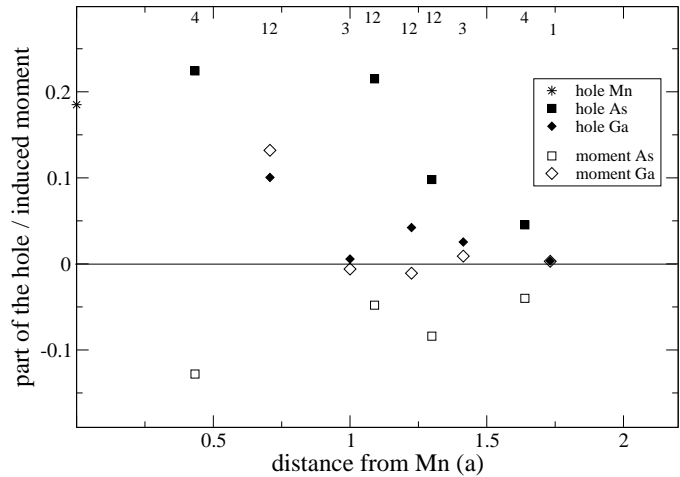}
\end{figure}

One of the important issues in the magnetism of the DMS is the spatial
localization of the hole states. Comparison of the DOS of pure GaAs and the GaMnAs with $x=3.125\%$
helps to get insight into the physical mechanism of the formation of the
hole states. The replacement
of one Ga atom in the supercell of GaAs by a Mn atom does not change the 
number of the spin-down states in the valence band. 
In the spin-up chanel there are, however, five additional
energy bands which are related to the Mn 3d states. 
Since there are five extra energy bands 
and only four extra electrons (the atomic configurations of Ga and Mn are 4s$^2$4p$^1$ and 3d$^5$4s$^2$) 
the valence band is not filled and there appear
unoccupied (hole) states at the top of the valence band. The integrated
number of the hole states is exactly one hole per Mn atom (correspondingly, one
hole per supercell).The distribution of the hole in the supercell for the
Mn concentration of 3.125\% is
shown in Fig. \ref{fig:hole_moment_distribution}. About 18.5\% 
of the hole is in the Mn sphere,
22.5\% 
in the first coordination sphere of the As atoms. Correspondingly 
about 60\% 
of the hole is outside of the first coordination sphere of the impurity.
About 58.3\% 
of the hole is on the As atoms, 17.7\% 
on Ga sites, 5.5\% 
in the empty spheres. Thus, the hole states are rather delocalized. Even the most distant As
atom contains 4.5\% 
of the hole. The delocalized character of the hole states is
an important factor in mediating the exchange interaction between
Mn atoms.

Another important quantity characterizing the localization of the
valence-band states about impurity is the values of the induced moments on various
atoms. The values of the atomic moments for $x=3.125\%$ are shown in 
Fig.\ref{fig:hole_moment_distribution}. It is seen
that even at the As atom most distant from Mn impurities
there is substantial spin polarization that provides an efficient exchange path
between Mn atoms. The dependence of the induced moment on the 
distance from Mn is nonmonotonous for both As and Ga. This behavior
can be related to the spin density oscillation in the Ruderman-Kittel-Kasuya-Yoshida 
(RKKY) theory.
The detailed analysis of the formation of the induced atomic moments
shows, however, that the physics here is more complex than the physics considered
by the RKKY theory since an essential role in the 
formation of the induced magnetic moments plays the hybridization of the 
As and Mn states. In particular, the negative sign of the moment 
of the As atoms is explained by the property that the empty (hole) states
have large As contribution. Since these states are of the spin-up type 
the spin-down As states become more occupied than the spin-up ones
leading to a negative induced moment. 

\section{Exchange parameters and Curie temperature}

The calculated exchange parameters are presented 
\begin{figure}
\caption{The parameters of the exchange interaction between Mn atoms (upper panel)
and the variation of $T_C^{MFA}$ with increasing number of the contributing 
coordination spheres (lower panel). The abscissa gives the radius of the coordination sphere
in the units of the lattice parameter of the zinc-blende crystal structure. 
\label{fig:exch_param}}
\includegraphics[width=8cm]{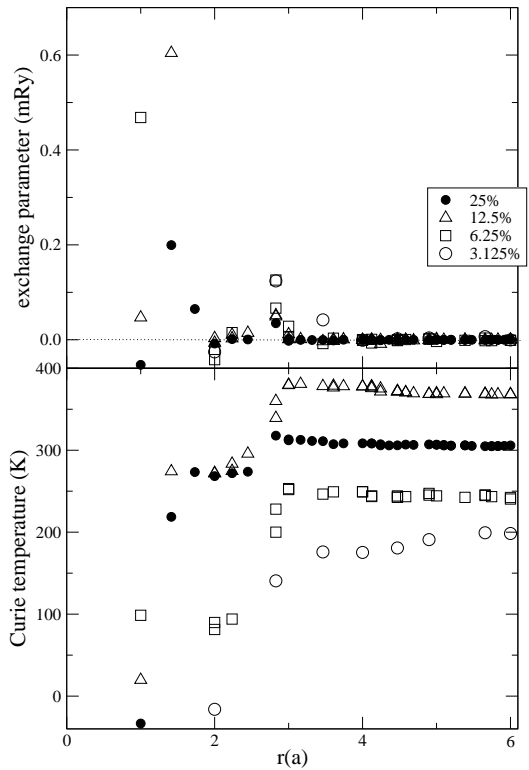}
\end{figure}
in Fig. \ref{fig:exch_param}. There are a number of conclusions that follow from the analysis 
of this figure. First, the Heisenberg model with the interaction
between the first nearest neighbors only
is not able to describe the magnetism of the system. For two Mn
concentrations ($x=25\%$ and $x=3.125\%$)
the first nearest-neighbor interaction is even antiferromagnetic, for $x=12.5\%$
it is positive but small. Second, the exchange interactions is rather quickly
decreasing with increasing distance between atoms.  
In Fig. \ref{fig:exch_param} we show the variation of the 
Curie temperature (\ref{eq:Tc_MFA}) 
with increasing number of the contributing coordination spheres. 
For instance, for concentrations $x=25\%,\:12.5\%,\:6.25\%$ no noticeable contribution to the
Curie temperature is obtained from the interactions between Mn atoms at the
distances larger than 3$a$. 
Third, the dependence of the exchange parameters on the distance between 
Mn atoms is not monotonous. At the distance of 2$a$ the interatomic interaction
is negative for all concentrations considered. Although the theory of the RKKY interaction
is not sufficient to describe the magnetism of the system, it is instructive to compare
the characteristic length of the variation of the calculated exchange parameters
with the period of the oscillation of the exchange parameters which follows
from the RKKY theory. Since the number of holes is exactly one per supercell
the characteristic volume of the Fermi sphere in a simple single-band
free-electron model is exactly the volume of the Brillouin Zone corresponding to the
given supercell. Thus for $x=25\%$ one can expect the oscillations with period 
close to $a$ and for $x=3.125\%$ with period close to $2a$. Indeed, the analysis
of the calculated exchange parameters shows that the first and second oscillation
of the exchange parameter for $x=25\%$ take place at the distance close to $a$
(from 1$a$ to 2$a$ and from 2$a$ to 3$a$). On the other hand the characteristic
range of the variation of the exchange parameter for $x=3.125\%$ is larger
(from 2$a$ to 4$a$). Thus there is a correlation between the number of holes and the 
characteristic range of the variation of the exchange parameters. 
\begin{figure} 
\caption{Curie temperature of Ga$_{1-x}$Mn$_{x}$As.
\label{fig:T_C}}
\includegraphics[width=8cm,angle=-90]{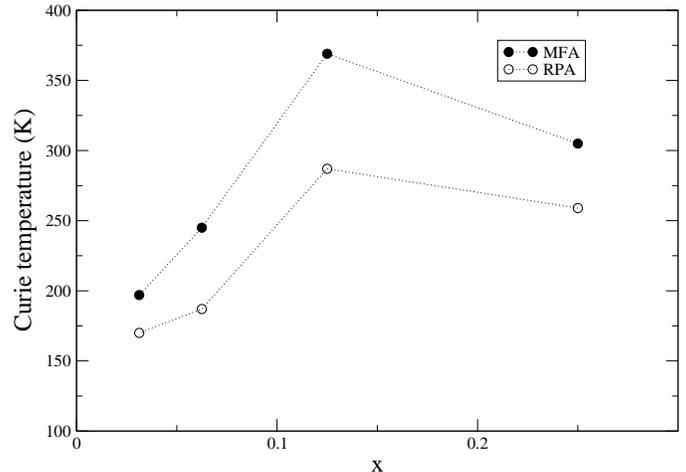}
\end{figure}

The calculated Curie temperatures are presented in Fig. \ref{fig:T_C}.
Both MFA and RPA approximations give similar dependence of the Curie temperature
on the Mn concentration: it increases at small $x$ has maximum at $x=12.5\%$ and
decreases for higher values of $x$. As already stated in Sect. \ref{sec:calc_scheme}, 
the 
$T_C^{MFA}$ is always larger than $T_C^{RPA}$. However, even $T_C^{RPA}$ 
exceeds substantially the experimental values of the transition temperature.
A possible explanation of this difference is the presence of the As antisites
and other donor defects that decrease the concentration of the holes in the
materials studied experimentally (see, e.g., Refs. \cite{akai,sanv_apl01,korz02}).
We address this issue in Sect. \ref{sect:T_C_of_holes}.

The nonmonotonous dependence of the transition temperature on the Mn concentration
is in qualitative agreement with experiment. \cite{matsukura,nonmon_comment} 
This nonmonotonous behavior is a consequence of the competition between different trends 
arising from the increase of the Mn 
concentration. On the one hand, the amplitude of the effective exchange
interaction between the Mn moments through the valence-band states of GaAs
increses with deacreasing distance between Mn atoms.  
Also the direct overlap of the Mn states increases. These
features produce a trend to the increase of the Curie temperature with 
increasing $x$.
On the other hand, an increased interaction
of the states of different Mn atoms results in the broadening
of the features of the partial Mn DOS at the Fermi level: 
As is clearly seen in Fig. \ref{fig:Mn_DOS}, 
at $x=3.125\%$ there is a narrow peak at the 
Fermi energy that is replaced by a broader structure with increasing $x$. 
Simultaneously the 
Mn magnetic moment decreases (Table~\ref{tab:moments}). These properties produce 
the trend to decreasing Curie temperature with increasing Mn concentration. 

Interestingly, the induced moment on the nearest As atoms
is the same for $25\%$  and  $12.5\%$ 
and decreases for $6.25\%$ and $3.125\%$ although
the inducing Mn moment is monotonously increasing (Table~\ref{tab:moments}). 
This shows that the nearest environment of 
each Mn atom is influenced by other Mn atoms even for the lowest Mn concentration.

\section{Dependence of the Curie temperature on the number of holes}
\label{sect:T_C_of_holes}

The experimental studies show that the concentration of holes in (Ga,Mn)As
is lower than the concentration of the Mn atoms. \cite{ohno_jmmm} 
One of the important factors leading to the low concentration of holes is
the presence of the As antisites ($\rm As_{Ga}$).
Since As has two more valence electrons compared with Ga, each As antisite
compensate the holes produced by two Mn atoms. 
 
\begin{figure}
\caption{$T_C^{MFA}$ for the (Ga,Mn)As with Mn concentration $x=0.0625$
as a function of the electron number $n$.
$n=0$ corresponds to the system  Ga$_{0.9375}$Mn$_{0.0625}$As with no
additional donor or acceptor defects.  
\label{fig:T_C_of_n}}
\includegraphics[width=8cm,angle=-90]{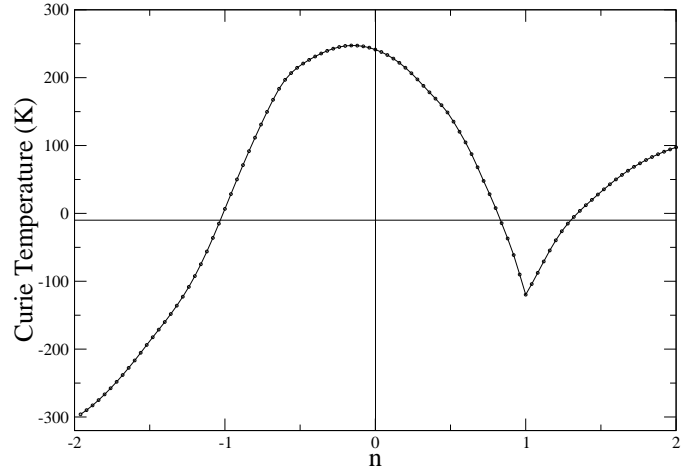}
\end{figure}

Here we use a simple rigid-band model to study the dependence
of the Curie temperature on the concentration
of antisites and other nonmagnetic donors. We assume that the 
electron structure calculated for (Ga,Mn)As with a given
Mn concentration will basically be preserved in the
presence of defects. The main difference
caused by the antisites concerns the occupation of the bands and, respectively, 
the position of the Fermi level.
In Fig.~\ref{fig:T_C_of_n} we show the dependence of the Curie
temperature calculated in the MFA as a function of the carrier number
for $x=6.25\%$. 
The calculated curve reveals strong dependence of the 
Curie temperature on the number of holes in the valence band. 
In agreement with the commonly
accepted picture of the hole-mediated ferromagnetism in (GaMn)As
the decrease of the number of holes leads to the decrease 
of the Curie temperature. 
At the number of additional electrons close to 0.85
(the relation of the hole concentration with respect to the Mn concentration
$\frac{p}{x}=0.15$) the ferromagnetic and antiferromagnetic interactions compensate
and the value of the mean field acting on the Mn spins in the ferromagnetic state 
becomes zero. Further decrease of the number of holes
leads to negative values of the mean field.
The negative value of the Curie temperature in Fig.\ref{fig:T_C_of_n}
means that in the assumed ferromagnetic configuration the antiferromagnetic
exchange interactions prevail resulting in a negative exchange field acting
on each Mn spin. The ferromagnetic state is unstable.

At $n=1$ the valence band is full. For larger $n$ the conduction band 
is occupied. 
Discontinuity in the character of the occupied states at $n=1$ results
in a kink in the $n$-dependence of the value of the mean field.
With occupation of the conduction band ($n>1$)
the derivative changes sign 
and the mean field increases with increase of $n$. For the number of electrons
in the conduction band larger than 0.28 per Mn atom the mean field becomes
positive. 
Remarkable is the asymmetry of the curve with respect to $n=1$. The curve is much
steeper to the left from the point than to the right from it. Correspondingly,
the Curie temperature at $n=2$ (one electron in the conduction band
per one Mn atom) is about 2.5 times smaller compared to the Curie temperature
at $n=0$ (one hole in the valence band per one Mn atom). This reveals 
higher efficency of the holes in the valence band in mediating
the exchange interaction between the Mn atoms compared with electrons
in the conduction band. The physical reason for this property is stronger
exchange interaction between the Mn moments and the states at the top of the 
valence band (p-d exchange) compared to the exchange interaction between
the Mn moments and the states at the bottom of the conduction band (s-d
exchange).

The region of negative $n$ (Fig.~\ref{fig:T_C_of_n}) corresponds to the concentration of 
of the holes higher than the concentration of the Mn atoms.
These states of the system can be obtained by avoiding the formation of $\rm As_{Ga}$ antisites
and by co-doping with atoms acting as donors. (Of course this simple theoretical consideration
does not take into account the 
difficulties in the producing of the materials with such properties.)

Moving in the direction of negative $n$, the value of the Curie temperature at first slightly 
increases. Then the mean field value decreases
fastly revealing again a trend to the antiferromagnetic exchange coupling.
The strong dependence of the mean field on the number of carriers reflects 
the competition between the antiferromagnetic and
ferromagnetic interactions. 

\section{Conclusions}

We use supercell and frozen-magnon approaches to study the dependence 
of the magnetic interactions in (Ga,Mn)As on the Mn concentration. We report 
the parameters of the exchange interaction between Mn spins and the 
MFA and RPA estimates of the Curie temperature. In agreement with
experiment we obtain a nonmonotonous dependence of the Curie temperature
on the Mn concentration. We estimate the dependence of the Curie temperature
on the concentration of the carries in the system and show that the decrease
of the number of holes in the valence band leads to fast decrease of the 
Curie temperature. The strong dependence of the Curie temperature on 
the carrier concentration provides an explanation to the overestimation
of the value of the Curie temperature compared to the experiment.

\begin{acknowledgments}
The financial support of Bundesministerium f\"ur Bildung und
Forschung is acknowledged. The authors are grateful to J. Kudrnovsky
and G. Bouzerar for interesting discussions and A. Ernst for sharing with us
his computer subroutines.
\end{acknowledgments}

 

\begin{thebibliography}{99}
\bibitem{nonmon_comment}
The maximum in the experimental dependence of the Curie temperature 
on the Mn concentration is at $x$ about 5$\%$ that is substantially smaller
that the theoretical value. A similar result was obtained by 
J. Kudrnovsky \cite{kudrn} within the CPA calculation. Our preliminary results whithin the approach
discussed in Sect. \ref{sect:T_C_of_holes} show that account for
As antisites shifts the position of the maximum to smaller $x$.
\end{thebibliography}
\end{document}